\documentstyle[amssymb,aps,twocolumn,epsf]{revtex}

\begin{document}
\draft
\title{A single atom in free space as a quantum aperture}
\author{S.J. van Enk and H.J. Kimble}
\address{Norman Bridge Laboratory of Physics, California\\
Institute of Technology 12-33, Pasadena, CA 91125}
\date{\today}
\maketitle

\begin{abstract}
We calculate exact 3-D solutions of Maxwell equations
corresponding to strongly focused light beams, and study their
interaction with a single atom in free space. We show how the
naive picture of the atom as an absorber with a size given by its
radiative cross section $\sigma =3\lambda ^{2}/2\pi $ must be
modified. The implications of these results for quantum
information processing capabilities of trapped atoms are
discussed.
\end{abstract}


The resonant absorption cross section for a single
two-state atom in free space driven by an electromagnetic
field of wavelength $\lambda$ is $
\sigma=3\lambda^{2}/2\pi$ \cite{jackson}. Thus it seems
reasonable to assume that a ``weak'' incident light beam
focused onto an area $A\sim $ $\sigma$ would experience a
loss (as resonance fluorescence) comparable to the incident
energy of the beam itself and would, for off-resonant
excitation, suffer an appreciable phase shift. In terms of
nonlinear properties, note that the saturation intensity
for a two-state atom in free space is $I_s=\hbar
\omega_a/2\sigma \tau$, where $\omega _{a}$ is the atomic
transition frequency and $\tau=1/\Gamma$ is the atomic
lifetime. Hence, a single-photon pulse of duration $T\sim $
$\tau$ should provide a saturating intensity and allow for
the possibility of nonlinear absorption and dispersion in a
{\it strong-focusing} geometry.

These considerations suggest that a single atom in free
space could perform important tasks relevant to quantum
information processing, such as nonlinear entangling
operations on single photons of different modes for the
implementation of quantum logic, along the lines of Ref.~
\cite{savage}, but now without the requirement of an
optical cavity \cite{quentin}. Further motivation on this
front comes from the need to address small quantum systems
individually, as for example in the ion-trap quantum
computer \cite {cz,wineland} or in quantum communication
protocols with trapped atoms in optical cavities
\cite{cirac,enk1}. Here, each ion (or atom) must be
individually addressed by focusing a laser beam with
resolution $\Delta x\gtrsim \lambda $ \cite{naegerl}.
Interesting effects may also be expected with respect to
the photon statistics of the scattered light in a regime of
strong focusing, such as extremely large photon bunching
\cite{car}. Conversely, alterations of atomic radiative
processes arising from excitation with squeezed and other
forms of nonclassical light would be feasible as well
\cite{gardiner}. Finally, questions of strong focusing
become relevant for dipole-force traps of size $\lambda $
for single atoms.

Against this backdrop of potential applications, we note
that radiative interactions of single atoms with strongly
focused light beams have received relatively little
attention. Indeed, previous experiments have been
restricted to a regime of {\it weak focusing} and
resultingly small fractional changes in transmission
\cite{ezekiel,itano,moerner}, either because of large focal
spot size $\sim 1000\lambda ^{2}$ \cite{itano} or reduced
oscillator strengths for molecular transitions
\cite{moerner}. On the theoretical front, we recall only
Refs.~\cite{car} studying the photon statistics by adopting
a quasi one-dimensional model.

In light of its fundamental importance, in this Letter we
report the first complete 3-dimensional calculations for
the interaction of strongly focused light beams and single
atoms in free space. Essential elements in this work are
exact 3-D vector solutions of Maxwell equations that
represent beams of light focused by a strong spherical
lens. As an application of our formalism, we calculate the
scattered intensities and the intensity correlation
function $g^{(2)}(0,\vec{r})$ as functions of angle for
resonant excitation of a single atom with a strongly
focused beam. We find an intriguing interplay between the
{\em angular} properties of the scattered light  and its
{\em quantum statistical} character (e.g., photon bunching
and antibunching versus scattering angle), leading to the
concept of a {\it quantum aperture}. Our results, in
particular those corresponding to scattering in the forward
direction, are compared to those of Ref.~ \cite{car}, and
to similar calculations using 3-dimensional paraxial
Gaussian beams \cite{siegman}, which we find do not always
represent the actual situation with strongly focused light
beams.

We start by constructing exact solutions of the Maxwell equations
describing tightly focused beams (a detailed analysis is deferred
to \cite{enku}, see also \cite{karman}). An incoming (paraxial)
beam with fixed circular polarization
$\vec{\epsilon}_{+}=(\hat{x}+i\hat{y})/\sqrt{2}$ and frequency
$\omega $ propagates in the positive $z$ direction and illuminates
an ideal lens. The incoming beam is taken to be a lowest-order
Gaussian beam with Rayleigh range $z_{{\rm in}}$ with $kz_{{\rm
in}}\gg 1$, and is characterized by the dimensionless amplitude
\begin{equation}
\vec{F}_{0}=\exp \left(-\frac{k\rho ^{2}}{2z_{{\rm
in}}}\right) \vec{\epsilon}_{+},
\end{equation}
where $\rho $ is the distance to the $z$ axis and the wave vector
$k=2\pi/\lambda$. For simplicity the focal plane of the incoming
beam and the plane of the lens are taken to coincide. After
transforming this input field through the lens, the output field
behind the lens is expanded in a complete set of modes
$\vec{F}_{\mu}$ that are exact solutions of the source-free
Maxwell equations adapted to the cylindrical symmetry of the
problem, as constructed in \cite{enk}. The index $\mu$ is
short-hand for the set of mode numbers $\mu=(k_t,m,s)$, with $k_t$
the transverse momentum number $ k_t=(k^2-k_z^2)^{1/2}$, $s$ the
polarization index and $m$ a topological index related to orbital
angular momentum \cite{enk}. For fixed $k$, the dimensionless mode
functions $\vec{F}_\mu$ are normalized to
\begin{equation}\int_{z={\rm const}}{\rm d} S
\vec{F}_\mu^*\cdot\vec{F}_{\nu}=
\delta(k_t-k_t')\delta_{mm'}\delta_{ss'} /(2\pi
k_t).\end{equation} As for the field transformation by the
lens, the action of a spherical lens is modeled by assuming
that the field distribution of the incoming field is
multiplied by a local phase factor $\exp (-ik\rho
^{2}/2f)$, with $f$ the focal length of the lens
\cite{goodman}. Thus, if in the plane of the lens, say
$z=0$, the incoming beam is given by $\vec{F}_{{\rm
in}}=\vec{F} _{0}$ as above, then the output field is given
by
\begin{equation}
\vec{F}_{{\rm out}}(\vec{r})=\int{\rm d}k_t \sum_m\sum_s
\kappa_\mu\vec{F}_{\mu }(\vec{r}), \label{out}
\end{equation}
where for the particular choice of $\vec{F}_{0}$, $\kappa_ \mu$ is
\cite{enku}
\begin{equation}
\kappa_\mu=\pi \delta
_{m1}\frac{k_{t}}{k}\frac{k_{z}+sk}{k} \xi \exp \left(
-\frac{k_{t}^{2}}{2k}\xi \right) ,
\end{equation}
with $\xi \equiv z_{R}-iz_{0}$, and
\begin{equation}
z_{R}=\frac{f^{2}z_{{\rm in}}}{z_{{\rm
in}}^{2}+f^{2}},\text{ \ }z_{0}=\frac{ fz_{{\rm
in}}^{2}}{z_{{\rm in}}^{2}+f^{2}}.
\end{equation}
In general, the expression (\ref{out}) for the outgoing
field must be evaluated numerically. In the paraxial limit,
when $kz_{R}\gg 1$, $z_{R}$ and $z_{0}$ correspond to the
Rayleigh range and the position of the focal plane of the
outgoing beam, respectively.

\begin{figure}[tbp]
\leavevmode    \epsfxsize=7cm \epsfbox{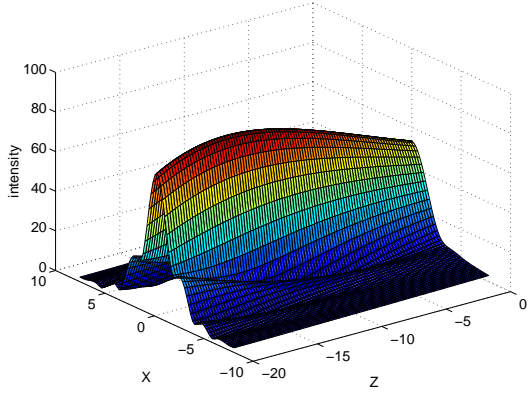}
\caption{Surface plot of the relative intensity $|\vec{F}_{{\rm
out}}\cdot\vec{\epsilon}_+|^2$ of a strongly focused beam as a
function of the dimensionless axial coordinate $Z\equiv (z-z_{0})/
\lambda $ and transverse coordinate $X\equiv x/\lambda $. The lens
is located at $z=0$ and is characterized by $f=500\lambda $, with
the incoming Gaussian beam having $z_{{\rm in}}/\lambda =6\times
10^{4}$. This implies $z_0\approx 500\lambda$ and $z_{R}\approx
4.2\lambda $. For paraxial beams, the focal plane would be at
$Z=0$.} \label{fig:surfc64.eps}
\end{figure}
A particular result for the $\vec{\epsilon}_{+}$ component
of $\vec{F}_{{\rm out}}(\vec{r})$ in the focal region is
given in Fig.~\ref {fig:surfc64.eps}. Note that the focal
plane deviates from the paraxial result $z=z_{0}$ and moves
towards the lens by several wavelengths. Furthermore, the
size of focal spot for the exact light beam is larger than
the corresponding value $\pi w_{R}^{2}$ with
$w_{R}\equiv\sqrt{z_{R}\lambda /\pi }$ for a paraxial beam.

With these results in hand, we now investigate the response
of an atom located at a position $\vec{r}_{0}$ in the focal
spot (i.e. the position of maximum field intensity) of a
strongly focused light beam as in Figure~1. The goal is to
identify the ``maximum'' effect that such an atom can have
on the transmitted and scattered fields. We consider a
$J_{g}=0\rightarrow J_{e}=1$ transition in the atom, as it
is the simplest case where all three polarization
components of the light in principle play a role. For the
cases presented here with the atom located on the $z$ axis,
the other two polarization components vanish \cite{enk},
but they can play a dominant role in other situations.

To calculate mean values of the scattered field as well as
its intensity and photon statistics, it is convenient to
work in the Heisenberg picture, in which the electric field
operator can be written as the sum of a ``free'' part and a
``source'' part, $\vec{E}=\vec{E}_{f}+\vec{E}_{s}$. The
source part for the case of a $J_{g}=0\rightarrow J_{e}=1$
transition is given by \cite{vogel}
\begin{equation}
\vec{E}_{s}^{(+)}(\vec{r})=\sum_{i}\vec{\Psi}_{i}(\vec{r}'
)\sigma _{i}^{-}(t-|\vec{r}' |/c).
\end{equation}
We have separated the fields into positive- and
negative-frequency components,
$\vec{E}_{f,s}=\vec{E}_{f,s}^{(+)}+\vec{E}_{f,s}^{(-)}$,
$\vec{r}' =\vec{r}-\vec{r}_{0}$, $\sigma _{i}^{-}$ is the
atomic lowering operator, and the sum is over three
independent polarization directions $ i=\pm 1,0$. In the
far field, $\vec{\Psi}_{i}(\vec{r})$ is the dipole field
\begin{equation}
\vec{\Psi}_{i}(\vec{r})=\frac{\omega _{a}^{2}}{4\pi
\varepsilon _{0}c^{2}} \left[
\frac{\vec{d}_{i}}{r}-\frac{(\vec{d}_{i}\cdot
\vec{r})\vec{r}}{r^{3}} \right] .
\end{equation}
Here $\vec{d}_{i}=d \hat{u}_{i}$ is the dipole moment
between the ground state $|g\rangle $ and excited state
$|e_{i}\rangle $ in terms of the unit circular vectors
$\hat{u}_{i}$ and the reduced dipole matrix element $d$.

Expressions containing the electric field in time-ordered and
normal-ordered form (as relevant to standard photo-detectors) can
be transformed into ``$ {\cal O}$-ordered form'', where
$\vec{E}_{s}^{(+)}$ is placed to the left of $\vec{E}_{f}^{(+)}$,
$\vec{E}_{f}^{(-)}$ to the left of $\vec{E}_{s}^{(-)}$, and where
the source parts are time-ordered\cite{vogel}. For instance, if we
assume the initial state of the field incident upon the lens is a
coherent state, then the normally ordered intensity
$I(t,\vec{r})\equiv \langle \vec{E}^{(-)}(t,\vec{r})\cdot
\vec{E}^{(+)}(t,\vec{r})\rangle $ can be written as the sum of
three terms, the intensity $I_{d}\equiv \langle
\vec{E}_s^{(-)}\cdot \vec{E}_s^{(+)}\rangle$ of the dipole field,
the intensity $I_{L}\equiv \langle \vec{E}_f^{(-)}\cdot
\vec{E}_f^{(+)}\rangle$ of the free (laser) field, and the
interference term between the two fields. Similarly, the
second-order correlation function $G^{(2)}(t,\tau ,\vec{r})\equiv
\sum_{l,m=x,y,z}\langle E_{l}^{(-)}(t)E_{m}^{(-)}(t+\tau
)E_{m}^{(+)}(t+\tau )E_{l}^{(-)}(t)\rangle $ (where the dependence
of the fields on $\vec{r}$ is suppressed) consists of 16 terms.
For $\tau =0$, $7$ of these vanish, yielding
\begin{eqnarray}
G^{(2)}(t,0,\vec{r}) =|\alpha |^{4}|\vec{F}_{{\rm out}
}|^{4}+2\sum_{i,j}|\alpha |^{2}|\vec{F}_{{\rm out}}|^{2}\vec{\Psi}
_{i}^{\ast }\cdot \vec{\Psi}_{j}\sigma _{ee}^{ij}(t_{r}) \nonumber
\\ +4\sum_{i}{\rm Re}(\alpha ^{\ast }\exp (i\omega
_{a}t))\vec{F}_{{\rm out}}^{\ast }\cdot \vec{\Psi}_{i}|\alpha
|^{2}|\vec{F}_{{\rm out}}|^{2}\sigma _{eg}^{i}(t_{r}))  \nonumber
\\ +2\sum_{i,j}|\alpha |^{2}(\vec{F}_{{\rm out}}\cdot
\vec{\Psi}_{i}^{\ast })( \vec{F}_{{\rm out}}^{\ast }\cdot
\vec{\Psi}_{j}\sigma _{ee}^{ij}(t_{r})),
\end{eqnarray}
where the coherent state amplitude is chosen such that
$\langle\vec{E}_f^{(-)}\rangle=\alpha \vec{F}_{{\rm out}}$, $t_r$
is the retarded time $t_{r}=t-|\vec{r}^{\prime }|/c$, and $\sigma
_{eg}^{i}=\langle \sigma _{i}^{-}\rangle $ and $\sigma
_{ee}^{ij}=\langle \sigma _{i}^{+}\sigma _{j}^{-}\rangle $ are
expectation values of the corresponding atomic operators.

To proceed beyond this point, we must evaluate the various atomic
quantities. As a simple starting point and in order to make
contact with the work of Ref. \cite{car}, we assume that the atom
reaches a stationary steady state. In this case, given the value
of the electric field at the atom's position  $\alpha\vec{F}_{{\rm
out}}(\vec{r}_0)$, the various atomic expectation values can be
straightforwardly derived \cite{vogel}.
\begin{figure}[tbp]
\leavevmode    \epsfxsize=7cm \epsfbox{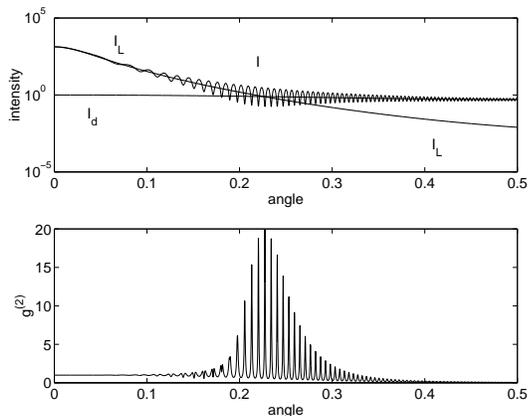} \caption{The
intensities $I_{L}$ of the laser (free) field, $I_{d}$ of the
dipole (source) field and $I$ of the total field relative to
$I_{d}(\protect\phi =0)$ as a function of the azimuthal angle
$\protect\phi /\protect\pi $, (i.e, at position $\vec{r}=[R\sin
\protect\phi ,0,R\cos \protect\phi ]$ where we chose
$R=50\protect\lambda $ here and for all further calculations). The
parameters for the incoming beam and the lens are as in Fig.~1,
and we chose $\lambda =852$nm, corresponding to the D2 transition
in Cs. The atomic dipole moment $ d$ is adjusted to give the
corresponding spontaneous emission rate $\Gamma =2\pi \times 5$MHz
for the 6$P_{3/2}$ states of Cs. (b) $g^{(2)}(0, \vec{r})$ as a
function of $\phi /\pi $. }
\end{figure}
For weak ($\alpha\rightarrow 0$) on-resonance excitation, we have
explicitly evaluated the scattered intensities as well as the
normalized second-order correlation function $g^{(2)}(\tau
,\vec{r})\equiv G^{(2)}(\tau ,\vec{r})/I^{2}(\vec{r})$ at $\tau
=0$ as functions of position in the far field. Recall that for a
stationary steady state, there is no dependence on $t$. As can be
seen in Figure~2a, in the forward direction (around $\phi =0$),
the free-field contribution $\vec{E}_{f}$ from the forward
propagating incident field overwhelms the source field
contribution $\vec{E}_{s}$ from the atom, even for focusing to a
spot of diameter $w_{R}\approx \lambda $ as in the figure (and in
fact is true for any width). This may be compared to a similar
result for classical scattering from spherical dielectrics with
light focused down to spot sizes larger than 5 times the size of
the spheres \cite{hodges}. Not surprisingly then, we find that
$g^{(2)}(0,\vec{r})\approx 1$ for forward scattering ($\phi \sim
0$) for {\em any} input beam, which, however, is in sharp contrast
with the result from \cite{car} which would predict a large
bunching effect (i.e., $g^{(2)}\gg 1$) for sufficiently tight
focusing. If we move instead to large angles ($\phi \sim \pi /2$),
Figure 2a shows that the dipole field $\vec{E} _{s}$ dominates
$\vec{E}_{f}$, so that $g^{(2)}(0,\vec{r})=0$ for $\phi
\rightarrow \pi /2$ (i.e., the light is almost purely fluorescence
and hence anti-bunched as for plane-wave excitation
\cite{antibunch}).

The behavior of $g^{(2)}$ is most interesting around the
angle $\phi _{0}$ where the incident $\vec{E}_{f}$ and
source $\vec{E}_{s}$ fields have the same magnitude.
Indeed, the oscillations apparent in Figure 2b indicate
that $g^{(2)}(0,\vec{r})$ is very sensitive to the relative
phase between $\vec{E} _{f}$ and $\vec{E}_{s}$. In fact,
maxima in $g^{(2)}$ appear when the free field and the
dipole field interfere destructively, which implies that
the total field is smaller than the contribution from the
incident field. Adopting the interpretation of Carmichael
and Kochan \cite{car} from an essentially one-dimensional
setting to the angular dependence of the fields around
$\phi _{0}$, we see that this implies that a photon has
just been absorbed by the atom, which is therefore in its
excited state, so that a fluorescent photon can be expected
to appear soon, thus leading to strong bunching. We suggest
that the combined angular dependences of $I(\vec{r})$ and
$g^{(2)}(0,\vec{r})$ evidenced in Figure 2 are
characteristic of scattering from a {\it quantum aperture}
such as an atom in free space.

\begin{figure}\label{f5p2} \leavevmode    \epsfxsize=7cm
\epsfbox{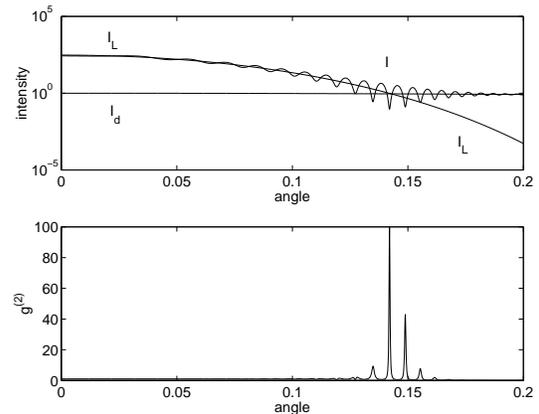}  \caption{ As Fig.~2, but for a
paraxial beam characterized by the same beam parameters
$z_R=4.2\lambda$ and $z_0=500\lambda$.}
\end{figure}
We have compared these exact 3-D results with those for a Gaussian
beam with the same parameters $z_{0}$ and $z_{R}$. In qualitative
terms, a Gaussian beam exaggerates the amount of light in the
forward direction at the cost of greatly underestimating it for
larger angles. This implies that the region where $g^{(2)}$
reaches its maximum is moved to smaller angles $\phi $ for a
paraxial beam as compared to the exact result (for the parameters
of Figure~2, $\phi _{0}\sim 26^{o}$ compared to $\phi _{0}\sim
40^{o}$, resp.). Moreover, the value of that maximum is
exaggerated as well, with a maximum value of
$g^{(2)}(0,\vec{r})\sim 100$ for the Gaussian beam. For even
stronger focusing, there will be large bunching at $\phi=0$ for a
paraxial beam, as in \cite{car}, but, as mentioned before, not for
the exact solutions.

Finally, we come back to the issue raised at the beginning
of this paper: why doesn't focusing a light beam to size
$\sigma $ give rise to large effects? One simple answer is
of course that there is a limit to how strongly one can
focus a light beam \cite{sales}, as indeed our exact
solutions show with focal areas $A$ always larger than
$\sigma $. Moreover, for tightly focused beams, the
polarization state in the focal volume is anything but
spatially uniform so that the field associated with a
single polarization for a paraxial input is split among
various components. In fact, if the atomic dipole is
$\vec{d}=d\hat{u}$, then the relevant quantity determining
the excitation probability is $|\hat{u}\cdot
\vec{E}^{(-)}(\vec{r }_{0})|^{2}$ evaluated at the atom's
position $\vec{r}_{0}$, while the total intensity in the
focal plane is given by $\int {\rm d}S|\vec{E}^{(-)}|^{2}$.
Thus, instead of $R= \sigma /A$, the scattering ratio
$R_{s}$ is
\begin{equation}
R_{s}=\frac{3\lambda ^{2}|\hat{u}\cdot
\vec{E}^{(-)}(\vec{r}_{0})|^{2}}{2\pi \int {\rm
d}S|\vec{E}^{(-)}|^{2}}.
\end{equation}
For a paraxial beam $R_s=2\sigma/(\pi w_R^2)\ll 1 $. For the lens
parameter used here, $f=500\lambda$, the optimum value (i.e.,
optimized over the parameters of the incoming beam) for $R_s$ is
$10\%$. Note that the ratio of the intensities of scattered
($\vec{E}_s$) and laser fields ($\vec{E}_f$) in the forward
direction ($\phi=0$ in Fig.~2) is much smaller than that (about
$10^{-3}$),  because the laser beam channels much more power in
that direction than does the dipole field.

For extreme values of $f$ on the order of $\lambda$, the maximum
scattering ratio does increase, but not beyond $50\%$. Even for a
scattering ratio of close to 50\% (reached for $f=2\lambda$ and
$z_{{\rm in}}=4\lambda$, for instance), the ratio of laser field
intensity $I_L$ to scattered intensity $I_d$ in the forward
direction ($\phi=0$) is not small, namely about 21.  Moreover, the
value for $g^{(2)}(\tau=0)=0.95$ agrees with Ref.~\cite{car} in
the sense that for the parameter $\Gamma\approx 0.5$ from that
paper antibunching is indeed predicted. On the other hand, it is
in contrast with the suggestion made there that very strong
bunching results for focusing to an area $A\sim \sigma$. Finally,
note that the upper limit of 50\% for $R_s$ can be understood by
noting that the optimum shape of the illuminating field would be a
dipole field. Here with light coming only from one direction, one
may indeed expect $R_{s}$ to be at most 0.5. With one mirror
behind the atom, an improvement in the scattering ratio by about a
factor of 2 might be expected. And of course, by building an
optical cavity around the atom, the atom-light interaction can be
further enhanced by orders of magnitude as in cavity quantum
electrodynamics \cite{cqed}.

In conclusion, by strongly focusing light on a single atom
in free space, one may indeed create an appreciable
light-atom interaction. However, this interaction is not as
strong as might be naively expected. On the one hand, this
implies that a coherent-state field employed for
``classical'' addressing of a single atom in
implementations of quantum computing and communication
\cite{cz,cirac,enk1} carries little information about that
atom, so that entanglement of the atom with other atoms in
a quantum register can be preserved \cite{enku}. On the
other hand, our analysis implies that there are serious
obstacles associated with using a single atom in free space
to process quantum information encoded in single photons.

We thank R. Legere for discussions. The work of HJK is supported
by the Division of Chemical Science, Office of Basic Energy
Science, Office of Energy, U.S. Department of Energy. SJvE is
funded by DARPA through the QUIC (Quantum Information and
Computing) program administered by the US Army Research Office,
the National Science Foundation, and the Office of Naval Research.

\end{document}